\documentstyle[12pt,aasms4]{article}
\tighten

\def\lsim{\lower.5ex\hbox{$\; \buildrel < \over \sim \;$}}
\def\gsim{\lower.5ex\hbox{$\; \buildrel > \over \sim \;$}}
\def\lax    {\ifmmode{_<\atop^{\sim}}\else{${_<\atop^{\sim}}$}\fi}
\def\gax    {\ifmmode{_>\atop^{\sim}}\else{${_>\atop^{\sim}}$}\fi}
\def\etal{{\it et al.\/} }
\def\gtorder{\mathrel{\raise.3ex\hbox{$>$}\mkern-14mu
             \lower0.6ex\hbox{$\sim$}}}
\def\ltorder{\mathrel{\raise.3ex\hbox{$<$}\mkern-14mu
             \lower0.6ex\hbox{$\sim$}}}

\def\pmb#1{\setbox0=\hbox{#1}%
  \kern-0.015em\copy0\kern-\wd0
  \kern0.03em\copy0\kern-\wd0
  \kern-0.015em\raise0.0433em\box0 }

\begin{document}

\title{ Kilohertz QPOs in Neutron Star Binaries
modeled as Keplerian Oscillations in a Rotating Frame of Reference}

\author{Vladimir Osherovich}
\affil{NASA/Goddard Space Flight Center/RITSS, Greenbelt MD 20771 USA;
vladimir@urap.gsfc.nasa.gov}

\author{Lev Titarchuk}
\affil{NASA/Goddard Space
Flight Center, Greenbelt MD 20771, and George Mason University/CSI, USA;
titarchuk@lheavx.gsfc.nasa.gov}

\vskip 0.5 truecm


\begin{abstract}

Since the discovery of kHz quasi-periodic oscillations (QPO) in neutron
star binaries, the difference between peak frequencies of
two modes in the upper part of the spectrum,  i.e.
$\Delta\omega=\omega_{h}-\omega_{k}$ has been studied extensively.
The idea that $\Delta\omega$  is constant and
(as a beat frequency) is related to the rotational frequency of the
neutron star has been tested previously.
The observed decrease of $\Delta\omega$ when $\omega_{h}$ and
$\omega_{k}$ increase has weakened the beat frequency
interpretation. We put forward a different paradigm: a Keplerian
oscillator  under the influence of the Coriolis force.
For such an oscillator, $\omega_{h}$ and the assumed Keplerian frequency
$\omega_{k}$ hold an upper hybrid frequency relation:
$\omega^2_{h}-\omega^2_{k}=4\Omega^2$, where $\Omega$ is the rotational
frequency of the star's magnetosphere near the equatorial plane.
For three sources (Sco X-1, 4U 1608-52 and 4U 1702-429), we demonstrate
that  the solid body rotation $\Omega=\Omega_0=const.$ is a good first
order  approximation. Within the second order approximation, the slow
variation  of $\Omega$ as a function of $\omega_K$ reveals the structure
of the  magnetospheric differential rotation. For Sco X-1, the QPO have
frequencies $\sim45$ and $90$ Hz which we interpret as the 1st and 2nd
harmonics of the lower  branch of the Keplerian oscillations for the rotator
with ${\bf \Omega}$ not aligned  with the normal of the disk:
$\omega_L/2\pi=(\Omega/\pi)(\omega_K/\omega_h)\sin{\delta}$ where
$\delta$ is the angle between ${\bf \Omega}$ and the vector normal to the
disk.

\end{abstract}

\keywords{accretion, accretion disks --- radiation mechanisms:
thermal ---stars:neutron --- X-rays: general}

\section{Introduction}

 Kilohertz quasi-periodic oscillations (KQPO) have been discovered
by the Rossi X-ray Timing Explorer (RXTE) in a number of
low mass X-ray binaries  (Strohmayer \etal 1996).
The existence of two observed peaks with frequencies $\omega_K$ and
$\omega_h$ in the upper part of the QPO spectrum
 became a natural starting point in modeling the phenomenon.
Attempts have been made to relate $\omega_K$, $\omega_h$ and the peak
difference frequency $\Delta\omega=\omega_h-\omega_K$ with the neutron
star spin and possible Keplerian motion of the hot matter surrounding the
star [see origin of Keplerian motion discussed in 
Titarchuk, Lapidus \& Muslimov 1998 (hereafter TLM) 
and Titarchuk \& Osherovich 1999].

In the beat frequency model, the burst oscillations and the kHz peak
separation $\Delta\omega$ are both considered to be close to the neutron
star spin frequency and thus $\Delta\omega$ is predicted to be constant
(see, e.g., review by van der Klis 1998). However, recent observations
of kHz QPO's in Sco X-1,  4U 1608-52 and 4U 1702-429 showed that
$\Delta\omega$ decreases systematically when $\omega_K$ and $\omega_h$
increases (van der Klis et al. 1997; Mendez et al. 1998;
 Markwardt, Strohmayer \& Swank 1999). Indications of such a decrease was
found by Wijnands et al. (1998) and Ford et al. (1998) for 4U 1735-44 as
well. Psaltis et al. (1998)  showed that the measured $\Delta\omega$ in
all other known sources is consistent with the values found in Sco X-1 and
4U 1608-52. These sources have become
standards for comparison between the kHz QPO phenomena in different
sources because of the high accuracy of measurements and the wide range of
values of $\omega_K$ and $\omega_h$ available.
For Sco X-1 and 4U 1608-52, $\Delta\omega/2\pi$ changes from
$320$ Hz to $220$ Hz when $\omega_K/2\pi$ changes from $500$ Hz to $850$ Hz

In the lower part of the Sco X-1 spectra, van der
Klis et al. (1997) found two frequencies $\sim45$ Hz
(referred to below as $\omega_L/2\pi$) and $90$ Hz which also slowly increase
 with increase of $\omega_K$ and $\omega_h$;
presumably 90 Hz is the second  harmonic of $\omega_L/2\pi$. Any
consistent model faces  a challenging task of describing the dependence
of $\Delta\omega$ and $\omega_L$ on $\omega_K$  and $\omega_h$ for
the available sources.

In the beat-frequency model (BFM) considered by Miller, Lamb \& Psaltis
(1998), the kHz QPO's are identified with the orbital frequency at the sonic
radius and its beat with the neutron star spin, and the lower frequency
 of horizontal branch oscillations (HBO) in Z-sources with the beat of
the orbital motion at the magnetospheric radius with the neutron star spin.
In the general-relativistic (GR) precession/apsidal  motion discussed
by Stella and Vietri (1998,1999), the kHz peaks are due to orbital motion and
GR apsidal motion of a slightly eccentric orbit, and low-frequency QPO to
the Lense-Thirring precession of the same orbit. 
Titarchuk and Muslimov (1997) suggested that a set of
peaks seen in the kHz QPOs can be naturally explained in terms of
the effect of rotational splitting of the main oscillation frequency in
the disk. The type of peak (eigenmodes) is dependent on 
mode numbers and a parameter describing disk structure.  
In TLM the authors argue that this effect of the rotational splitting 
originated  in a disk region identified as the
centrifugal barrier (CB) region oscilatted in the vertical and radial 
directions.  The size oscillations of the CB disk region
and reprocessing of the disk photons there (due to Comptonization) led to an 
oscillation signal.   

In this {\it Letter}, we take a different approach and 
we suggest that a kHz QPO can be explained as oscillations of large
scale inhomogeneities (hot blobs) thrown into the neutron star's
magnetosphere. Participating in the radial oscillations with Keplerian
frequency $\omega_K$, such blobs are simultaneously under the influence of
the Coriolis force. The present model is very different than the TLM one 
where the site of the rotational splitting was associated with the CB disk
region. Thus, studying the Keplerian oscillator in the rotating
frame of reference (magnetosphere), we attempt to relate $\Delta\omega$ 
and $\omega_L$ with $\omega_K$ and $\omega_h$. In our model, we compare 
results of linear theory
with the observations for three sources. We stress the fundamental role
of the Keplerian frequency (which we identify with $\omega_K$)
and the upper hybrid frequency relation between $\omega_K$ and
$\omega_h$. In this first {\it Letter}, we depict the main features of
the kHz QPO phenomena. The discussion identifies some limitations
of our approach and suggests further directions of the research we shall
pursue.

\section{Keplerian Oscillator Under the Influence of the Coriolis Force}

We assume that the interplay
between the centrifugal, gravitational force  and magnetic force 
  maintains radial oscillations (along the axis x) with
the Keplerian frequency
\begin{equation}
\omega_K=\left(\frac{GM}{R^3}\right)^{1/2}
\label{two}
\end{equation}
where G is the gravitational constant, M is the mass of the compact
object and R is the radius of an orbit. 
Self-similar spheromak-type models of a magnetic atmosphere 
 can serve as an example of the exact time dependent 
MHD solutions in which non-radial equilibrium is maintained by 
magnetic force, while all motion is strictly radial 
(Farrugia et al. 1995 and references there in). Taking the magnetospheric 
rotation with  angular velocity $\bf \Omega$ (not perpendicular to 
the plane of the equatorial disk), 
the small amplitude oscillations of such a blob thrown into the magnetosphere
 are described  by  equations (Landau and Lifshitz, 1960)  
\begin{equation}
\ddot{x}+\omega_K^2x-2\dot{y}\Omega\cos\delta=0
\label{three}
\end{equation}
\begin{equation}
\ddot{y}+2\dot{x}\Omega\cos\delta-2\dot{z}\Omega\sin\delta=0
\label{four}
\end{equation}
\begin{equation}
\ddot{z}+2\dot{y}\Omega\sin\delta=0
\label{five}
\end{equation}
where $\delta$ is the angle between $\bf\Omega$ and the vector normal to
the plane of radial oscillations, y is the azimuthal component of the 
displacement vector and z is the vertical component which is perpendicular 
to the plane of Keplerian oscillations. The dispersion relation
for the frequency $\omega$ 
\begin{equation}
\omega^2[\omega^4-(\omega_K^2+4\Omega^2)\omega^2+4\Omega^2\omega_K^2\sin^2\delta
]=0
\label{six}
\end{equation}
 besides the non-oscillating mode ($\omega=0$), describes
two spectral branches (eigenmodes), high  and low:
\begin{equation}
(\omega^2)_{1,2}=\frac{\omega_h^2\pm(\omega_h^4-16\Omega^2\omega_K^2\sin^2\delta
)^{1/2}}{2}
\label{seven}
\end{equation}
where
\begin{equation}
\omega_h=(\omega_K^2+4\Omega^2)^{1/2}
\label{eight}
\end{equation}
is the analog of the upper hybrid frequency $f_{uh}$ in plasma physics
(Akhiezer et al., 1975; Benson 1977). When the angle $\delta$ is small 
(as we believe is
indeed our case):
\begin{equation}
\omega_1=\omega_h
\label{nine}
\end{equation}
is related with the  radial eigenmode and 
\begin{equation}
\omega_2\equiv\omega_L=2\Omega(\omega_K/\omega_h)\sin \delta
\label{ten}
\end{equation}
is related with the vertical eigenmode.
For $\delta\ll 1$, corrections for $\omega_1$ are of the second order in
$\delta$, but for  $\omega_2$, the angle dependent term is the main term.

\section{Comparison of the Model with Observations}
According to formula (7)
\begin{equation}
\Omega=\frac{(\omega_h^2-\omega_K^2)^{1/2}}{2}
\label{eleven}
\end{equation}
is the rotational frequency, which, for the highly conductive plasma in
the case of corotation, should be approximately constant ($\Omega=\Omega_0$).
Interpreting the two frequencies in the upper part of the kHz spectrum as
$\omega_K$ and $\omega_h$ [where $\omega_K<\omega_h$] for Sco X-1, we plotted
in Figure 1, $\nu=\Omega/2\pi$ versus $\nu_K=\omega_k/2\pi$ (solid circles).
Indeed, one may notice that as a first approximation $\Omega=\Omega_0=const.$
The slow (but systematic) variation of $\Omega/2\pi$ is
between 330 and 350 Hz which should be compared with the variation by a
factor of 1.5 for $\Delta\omega=\omega_h-\omega_K$ for the same range
of $\omega_K$. If the magnetosphere corotates with the neutron star
(solid body rotation), then our procedure would determine
the spin rotation of the star. In fact, there is a differential rotation
profile which depends on the magnetic field $\bf{B}$  which can be presented 
in terms of the
Chandrasekhar potential $A$ (Chandrasekhar, 1956)
\begin{equation}
{\bf{B}} =\frac{1}{R}\left(-\frac{1}{R}\frac{\partial A}{\partial \mu},
-\frac{1}{(1-\mu^2)^{1/2}}\frac{\partial A}{\partial
R},\frac{B^*(A)}{(1-\mu^2)^{1/2}}\right)
\label{twelve}
\end{equation}
where $\mu=\cos\theta$, $\theta$ is the colatitude in the spherical
system of coordinates and $B^*(A)$ is a function of $A$ only.
For a combination of dipole, quadrupole and octupole fields
\begin{equation}
A=A_0(1-\mu^2)\left[\frac{1}{R}+\frac{q}{R^2}\mu-\frac{b}{R^3}(5\mu^2
+2)\right]
\label{thirteen}
\end{equation}
where $A_0$, $q$ and $b$ are constants. For a highly conductive plasma,
the MHD induction equation leads to the iso-rotation theorem of Ferraro
(see Alfv\'en and Falthammar, 1963) which in our notation means that
$\Omega$ is a function of $A$ only. Following a recent suggestion of
Osherovich and Gliner (1999), we assume that in the vicinity of a magnetic
star $\Omega(A)$ has a maximum and therefore can be presented as
\begin{equation}
\Omega(A)=\Omega_0-a^2A^2
\label{fourteen}
\end{equation}
where $\Omega_0$ and $a$ are constants. Since in the
equatorial plane ($\mu=0$), according to (12), the quadrupole term does not
contribute to $A$, from formula (13) we find
\begin{equation}
\Omega(A)=\Omega_0-\left(\frac{\alpha^{\prime}}{R}-
\frac{\beta^{\prime}}{R^3}\right)^2
\label{fifteen}
\end{equation}
where $\alpha^{\prime}$ and $\beta^{\prime}$ are positive constants.
Expressing $R$ through the Keplerian frequency
$\nu_K=\omega_K/(2\pi)$ from equation (1), we find that formula (14)
leads to the following presentation for
$\Omega$:
\begin{equation}
\Omega(\nu_K)/2\pi=C_0+C_1\nu_K^{4/3}+C_2\nu_K^{8/3}+C_3\nu_K^4
\label{sixteen}
\end{equation}
where $C_2=2\sqrt{C_1C_3}$ (not a free parameter because
the radial part of the expansion is a full square of $\nu_K^{2/3}$ and 
$\nu_K^{2}$ with new positive constants $\alpha$ and $\beta$).
The fit for Sco X-1 (solid line in Figure 1) by formula (16)
demonstrates the agreement of the proposed model with the data. Indeed,
for Sco X-1, we find $C_0=\Omega_0/2\pi=345$ Hz,
$C_1=-\alpha^2=-3.29\cdot10^{-2}$ Hz$^{-1/3}$,
$C_2=2\alpha\beta=1.017\cdot10^{-5}$ Hz$^{-5/3}$ and
$C_3=-\beta^2=-7.76\cdot10^{-10}$ Hz$^{-3}$
with high $\chi^2=37.6/39$.

 In our fitting procedure, $C_0$, $C_1$,
$C_2$ and $C_3$ are treated as four free parameters, i.e.,
the relation between $C_2$ and the two constants $C_1$
and $C_3$ is not enforced. The fact that the derived $C_2$ indeed
satisfies the relation $C_2=2\sqrt{C_1C_3}$ shows the internal consistency of
the model.

Three available data points for the 4U 1702-429 source are consistent with
a constant value of rotation frequency
$\Omega_0/2\pi=380$ Hz, but the small number of data points does not
allow us to recover the profile of $\Omega(\nu_K)$. However, the fit
of 4U 1608-52, shown in Figure 2, demonstrates that the
excellent agreement between our model and observations for Sco X-1 is
not accidental. For 4U 1608-52, we found $C_0=\Omega_0/2\pi=345$ Hz,
$C_1=-3.91\cdot10^{-2}$ Hz$^{1/3}$, $C_2=1.23\cdot10^{-5}$ Hz$^{-5/3}$ and
$C_3=-9.45\cdot 10^{-10}$ Hz$^{-3}$ with $\chi^2=12.02/13$.

Sco X-1 allows us to check the prediction of the model for
$\omega_L$ [formula (10)]. Taking the profile $\Omega(\nu_K)$, presented
in Figure 1, we plot $\nu_L=\omega_L/2\pi$ in Figure 3. Choosing
$\delta=5.5^o$, we found that our plot for $\nu_L$ and $2\nu_L$ fit the
data for the observed frequencies of $~45$ Hz and $~90$ Hz, respectively.
From these results, we conclude that the rotational axis of
Sco X-1 and the normal to the plane of radial oscillations do not
coincide but are separated by $5.5^o$.
\section{Summary and Discussion}
The main thrust of this work is based on the hypothesis that two of the
observed frequencies in the upper part of the kHz QPO spectrum are the
Keplerian frequency ($\omega_K$) and the upper
hybrid frequency ($\omega_h$) of the Keplerian oscillator in the frame
of reference rotating with the angular frequency $\Omega$.
For three sources, we showed that $\Omega$ calculated
according to the formula $\Omega=\sqrt{\omega_h^2-\omega_K^2}/2$ changes
with $\omega_K$ significantly less than $\Delta\omega=\omega_h-\omega_K$,
suggesting approximately constant rotation of the star's magnetosphere.
The detailed profile of $\Omega$ is successfully modeled
as a function of $\nu_K$ within the dipole-quadrupole-octupole
approximation of the magnetic field for Sco X-1 and for 4U 1608-52.
For Sco X-1, the derived $\Omega(\nu_K)$ is used to compare
predictions of the model for the lower branch frequency $\omega_L/2\pi$
with observations of $~45$ Hz and $90$ Hz (presumably $2\omega_L/2\pi$).
The last fit is done with only one unknown parameter
$\delta$ (angle between the $\bf\Omega$ and the normal vector to the
plane of Keplerian oscillations).

A good fit for $\omega_L$ and $2\omega_L$
confirmed the suggested model and
revealed the expectedly small angle $\delta=5.5^o$. The profile for
$\Omega(\nu_K)$ has been modeled strictly in the equatorial plane ($\mu=0$)
for the ideally aligned rotator ($\delta=0$). With $\delta\neq0$,
the next approximation will reveal the asymmetry of the QPO
spectrum due to the quadrupole term. It has been shown (Osherovich, Tzur
and Gliner, 1984) for the solar corona that the quadrupole term introduces
North-South asymmetry in the makeup of the magnetic atmosphere surrounding a
gravitating object. We shall pursue a similar study for the
neutron star with the expectation of finding a signature of such
asymmetry in the kHz QPO.
Recent measurements show that the octupole field is comparable to the
global dipole field near the sun, while the quadrupole field contributes
$20-30\%$ of the total field (Wang et al., 1997;
Osherovich et al., 1999).

In this {\it{Letter}}, we restrict the scope
of the work to a model for the linear Keplerian oscillator.
The presence of the second harmonic ($~90$ Hz) of the $~45$ Hz
mode, noticed by van der Klis et al. (1997 ), strongly suggests that a
weakly nonlinear model is desirable. 
 Some oscillations below $\omega_L$ (for Sco X-1
oscillations with frequencies $10-20$ Hz) we attribute to the innermost
edge of the Keplerian disk, which adjusts itself to the rotating central
object (i.e., neutron star). The physics of
these oscillations related to accreting matter with a large angular
momentum is outside the scope of this Letter and it have been described in
a separate paper (Titarchuk \& Osherovich 1999).

We are grateful to J. Fainberg and R.F. Benson for help in the preparation
of this paper and discussions and for fruitful suggestions by the referee.
L.T. thanks NASA for support under grants  NAS-5-32484 and  
RXTE Guest Observing Program and Jean Swank for support of this work. 
In particular, we are grateful to Michiel van der Klis and  Mariano Mendez 
for the data which enabled us to make detailed
comparisons of results of  our model with the data.

\clearpage

\begin{figure}
\caption{The inferred rotational frequency of the NS magnetosphere 
(see Eq. 10)  as a function of the frequency of the low
kHz QPO peak (Keplerian frequency).
 For Sco X-1 (van der Klis et al. 1997) (solid circles) and
for 4U 1702-429 (Markwardt et al. 1999) (open squares).
The solid (in the case of Sco X-1) line and dashed line (in the case of
4U 1702-429) are  theoretical curves calculated using the multipole
expansion of the rotational frequency (Eq.15 ).
\label{Fig1}}
\end{figure}

\begin{figure}
\caption{The inferred rotational frequency of the NS magnetosphere 
(see Eq. 10) as a function of the frequency of the low kHz QPO peak (Keplerian
frequency). For  for 4U 1608 (Mendez et al. 1998) (solid squares).
The solid line  is the theoretical curve (see Eq.15).
\label{Fig2}}
\end{figure}

\begin{figure}
\caption{HBO frequency vs high peak KQPO frequency.
for Sco X-1 (van der Klis et al. 1997) (solid circles).
The solid  line is  theoretical dependence  of the low branch oscillation
frequencies (40-50 Hz) and second harmonics (80-100 Hz) on the
high hybrid frequencies, calculated using the multipole
expansion of the rotational frequency (Eq. 16) with $\delta=5.5^o$..
\label{Fig.3}}
\end{figure}

\noindent

\begin{thebibliography}{}

\bibitem[Akhiezer et al. (1975)]{A75} Akhiezer, A.I., Akhiezer, A.I.,
Polovin, R.V.,
Sitenko, A.G. \&  Stepanov, K.N. 1975,  Plasma Electrodynamics   (New
York: Pergamon)

\bibitem[Alfv\'en \& Falthammar (1963)]{afl63} Alfv\'en, H., \&
Falthammar, C-G. 1963,
Cosmical Electrodynamics.  Fundamental Principles (London: Qxford
University Press)

\bibitem[Benson \etal  ]{benson} Benson, R.F. 1977, Rad. Sci., 12, 861

\bibitem[Chandrasekhar, 1956]{chand}
Chandrasekhar, S. 1956, Proc. Acad. Sci. 42, 1

\bibitem[Farrugia \etal (1995)]{far95}
Farrugia, C.J., Osherovich, V.A. \& Burlaga, L.F.   1995, J. Geophys. Res.,
 100,  293

\bibitem[Ford \& van der Klis (1998)]{fvk}
Ford, E.,  van der Klis, M., 
van Paradijs, J., Wijnands, R., Mendez, M., \& Kaaret, P.
 1998, \apj , 508, L155




\bibitem[Landau \& Lifshitz (1960)]{ll60}
Landau, L.,  \& Lifshitz, E. 1960,  Mechanics
(New York: Pergamon)

\bibitem[Markwardt  \etal (1999)]{mss}
Markwardt, C.B., Strohmayer, T.E., \& Swank, J.H.
1999, \apj, in press

\bibitem[Mendez  \etal (1998)]{men98}
Mendez, M., van der Klis, M., Ford, E.C., Wijnands, R.,
van Paradijs, J., \& Vaughan, B.A.  1998, \apj , 505, L23

\bibitem[Miller  \etal (1998)]{mlp}
Miller, M.C.,  Lamb, F.K. \& Psaltis, D.
1998, \apj,  508, 791

\bibitem[Osherovich \etal (1984)]{osh}
Osherovich, V.A., Tzur, I., \& Gliner, E.B. 1984, \apj, 284, 412

\bibitem[Osherovich \etal (1999)]{osh}
Osherovich, V.A., \& Gliner, E.B. 1999, in preparation

\bibitem[Psaltis \etal (1998)]{ps98}
Psaltis, D., et al. 1998, \apj , 501, L95

\bibitem[Swank  (1998)]{sw98}
Swank J.H., 1998, IAU SYMP. 188, 107

\bibitem[Shakura \& Sunyaev (1973)]{SS73}
Shakura, N.~I. \& Sunyaev, R.~A. 1973, A\&A, 24, 337

\bibitem[Stella,  \&  Vietri (1998)]{SV}
Stella, L., \& Vietri, M, 1998, \apj, 492, L59

\bibitem[Stella,  \&  Vietri (1999)]{SV}
Stella, L., \& Vietri, M. 1999, Phys. Rev. Letters, 82, 17

\bibitem[Strohmayer \etal (1996)]{str96a}
Strohmayer, T.~E., Zhang, W., Swank, J.~H., Smale, A.,
Titarchuk, L., Day, C., \&  Lee, U. 1996, \apj , 469, L9

\bibitem[Swank (1998)]{tm98}
Swank, J., astro-ph/9802188 to appear in
 ``The active X-ray Sky'' eds. L. Scarsi, H Bradt, P. Giommi, and F.
Fiore   (1998)

\bibitem[Titarchuk, Lapidus \& Muslimov (1998)]{tlm98}
Titarchuk, L., Lapidus, I.I.,\& Muslimov, A. 1998, \apj, 499, 315 

\bibitem[Titarchuk \& Osherovich (1998)]{to99}
Titarchuk, L.,\& Osherovich, V. 1999, \apj, 518, in press 

\bibitem[Titarchuk \& Muslimov (1997)]{tm97}
Titarchuk, L., \& Muslimov, A. 1997, \aap, 323, L5

\bibitem[van der Klis  1998]{vk98}
van der Klis, M. 1998, in AIP Conf Proc 431, 361


\bibitem[van der Klis  1997]{vk97}
van der Klis, M., Wijnands, R.A.D., Horne, K. \& Chen, W.
1997, ApJ, 481, L97



\bibitem[wijnands  1997]{w97}
Wijnands, R.A.D., \& van der Klis, M. 1997, Nature, 394, 344

\bibitem[Wang  \etal 1997]{W97}
Wang, Y-M., \etal 1997, ApJ, 485, 875


\end{thebibliography}
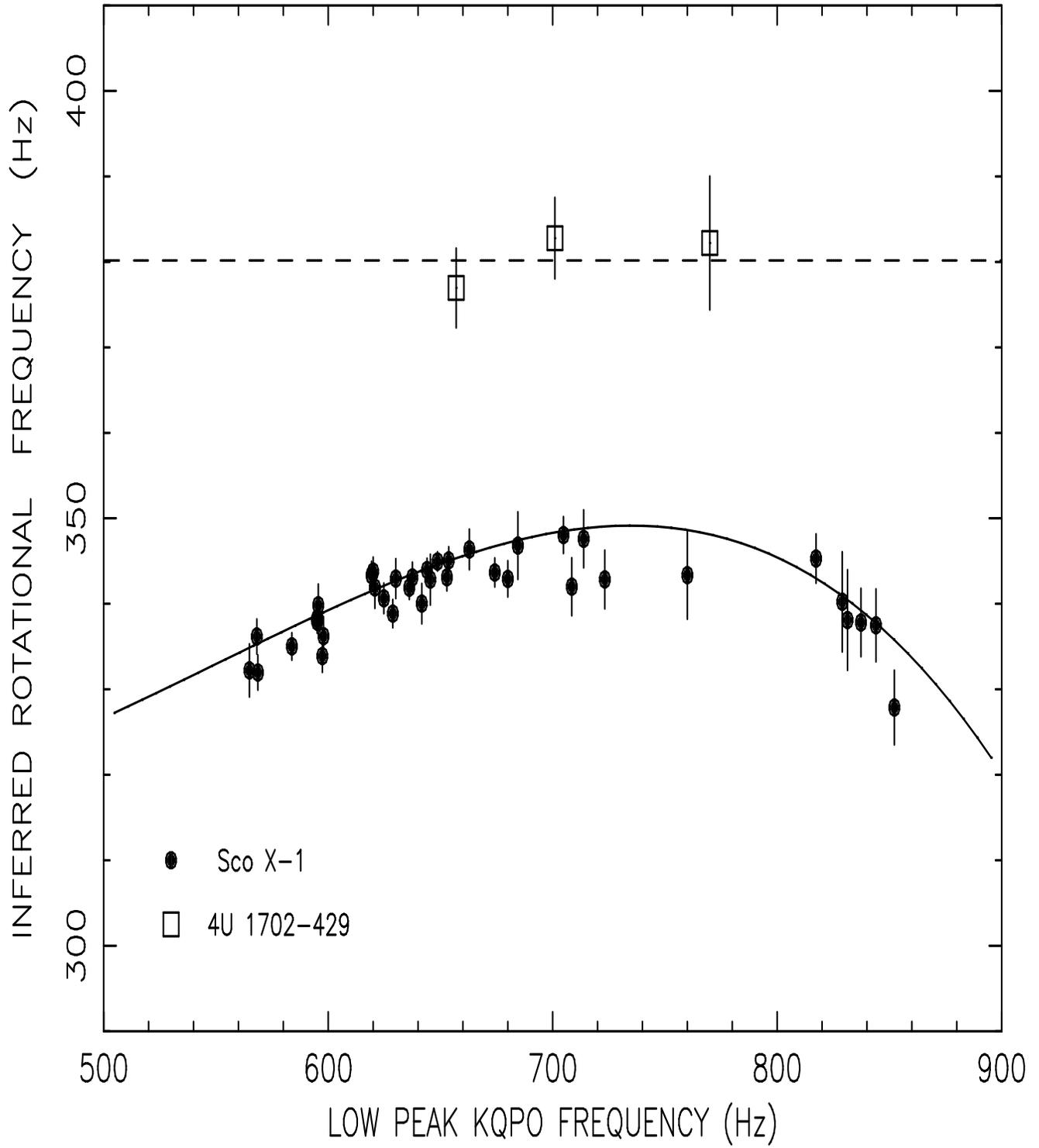
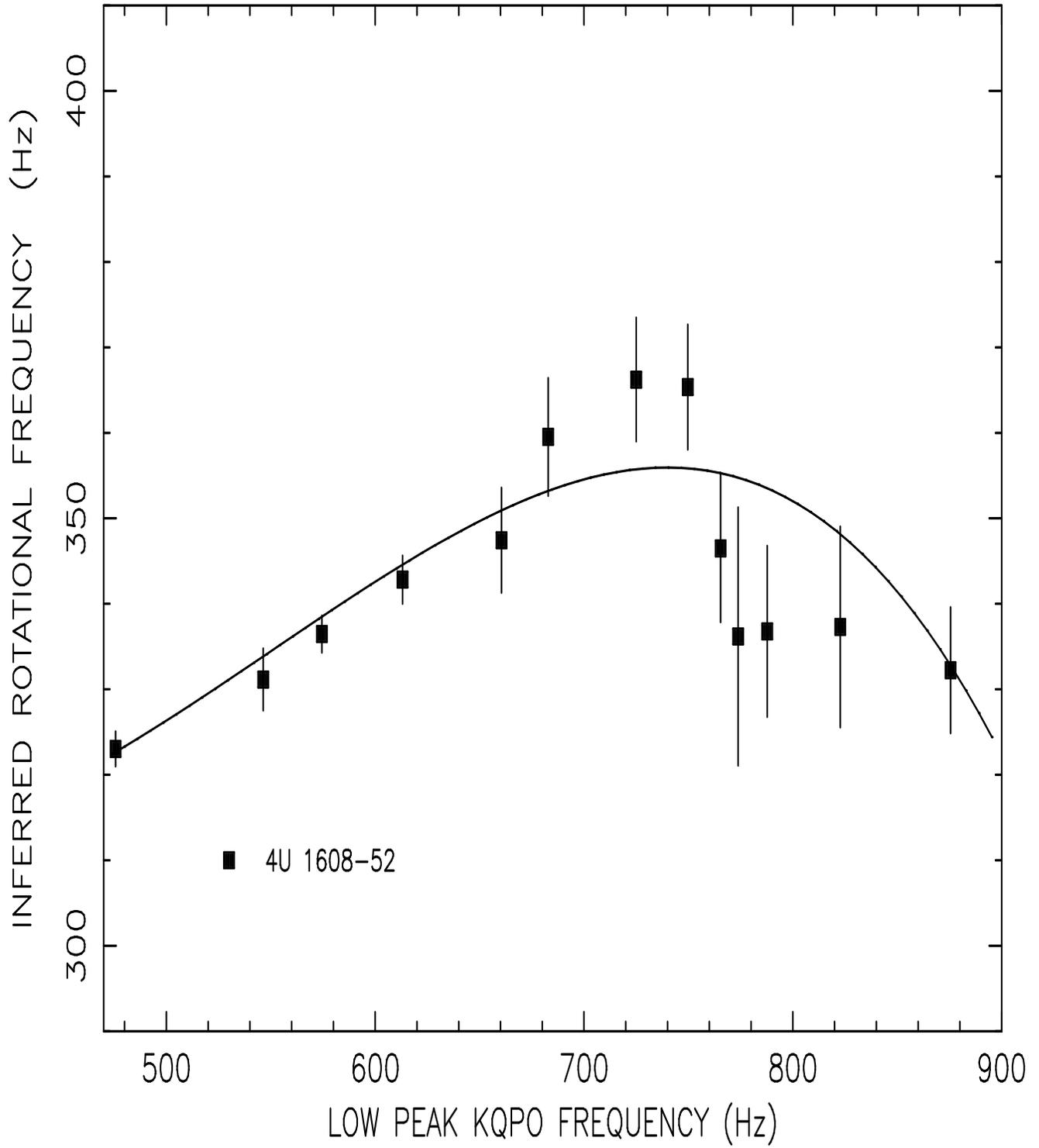
\end{document}